\documentstyle[aaspp4,psfig,epsfig]{article}

\begin{document}

\title{XTE~J2123--058:  A New Neutron Star X-Ray Transient}

\authoremail{jat@astro.columbia.edu}

\author{John A. Tomsick, Jules P. Halpern, Jonathan Kemp\altaffilmark{1,2}}
\affil{Department of Physics and Columbia Astrophysics Laboratory, Columbia University, 550 West 120th Street, New York, NY 10027 (e-mail: jat@astro.columbia.edu)}
\altaffiltext{1}{Visiting Astronomer, Cerro Tololo Inter-American Observatory, National
Optical Astronomy Observatories, which is operated by the Association of
Universities for Research in Astronomy, Inc.~(AURA) under cooperative agreement
with the National Science Foundation.}
\altaffiltext{2}{Visiting Astronomer, Kitt Peak National Observatory, National Optical
Astronomy Observatories.}

\author{Philip Kaaret}
\affil{Harvard-Smithsonian Center for Astrophysics, 60 Garden Street, Cambridge, MA 02139 (e-mail: pkaaret@cfa.harvard.edu)}

\begin{abstract}

We report on optical and RXTE observations of a new high-latitude bursting X-ray
transient, XTE~J2123--058.  We identified the optical counterpart and discovered
a $5.9573\pm 0.0016$~hr periodic optical modulation, which was subsequently shown 
to be the same as the spectroscopic orbital period.  From the absence of orbital 
X-ray modulation and the presence of partial optical eclipses the binary 
inclination is between $55^{\circ}$ and $73^{\circ}$.  From the optical magnitude 
in quiescence and from the X-ray flux of type I X-ray bursts, the source 
distance is between 4.5 and 15~kpc, which implies that the source is unusually 
far from the Galactic plane since $b = -36.2^{\circ}$.  Optical bursts with
properties consistent with being reprocessed X-ray bursts occurred.

We detect a pair of high frequency QPOs at $847.1\pm 5.5$~Hz and $1102\pm 13$~Hz
simultaneously.  According to the beat frequency model, this QPO separation 
implies a neutron star spin period of $3.92\pm 0.22$~ms.  A change in the energy 
spectrum occurred during the decay of the outburst, which may have been due to the 
onset of the propeller mechanism.  If so, then the neutron star magnetic field 
strength is between 2 and 8~$\times 10^{8}$~G for an assumed distance of 10~kpc.
However, the changes in the timing and spectral properties observed during the 
decay are typical of atoll sources, which may indicate that the changes are
due solely to the dynamics of the accretion disk.

As the phase averaged V-band magnitude declined from 17.26 at the peak of the 
outburst to 19.24, and the X-ray flux decreased from 
$9.6\times 10^{-10}$~erg~cm$^{-2}$~s$^{-1}$ to $7.3\times 10^{-11}$~erg~cm$^{-2}$~s$^{-1}$,
the peak-to-peak amplitude of the V-band modulation increased from 0.75 to 1.49 
magnitudes.  This behavior can be explained if the size of the accretion disk 
decreases during the decay of the outburst.

\end{abstract}

\keywords{accretion, accretion disks --- X-ray transients: general ---
stars: individual (XTE~J2123-058) --- stars: neutron stars --- X-rays: stars}

\clearpage

\section{Introduction}

The X-ray transient XTE~J2123--058 was discovered by the \it Rossi X-Ray
Timing Explorer \rm All-Sky Monitor (RXTE/ASM) on June 27, 1998 
(\cite{levine98}).  X-ray bursts were detected in pointed RXTE observations 
made on June 27 and June 29 (\cite{takeshima98}), indicating that the system 
contains a neutron star.  Based on the burst profiles and time intervals
between bursts, we classify the bursts as type I (\cite{lewin95}).  If this 
system is a neutron star low mass X-ray binary (LMXB), as the data suggest, 
it is unusual because of its high Galactic latitude ($b = -36.2^{\circ}$).

In this paper, we report on optical and RXTE observations of XTE~J2123--058.
Our results include the optical identification of XTE~J2123--058 (\cite{tomsick98a}), 
the measurement of a 5.9573~hr optical photometric modulation during its outburst 
(\cite{tomsick98b}), which proved to be the binary orbital period (\cite{hynes98}), 
and constraints on the binary inclination.  We place a lower limit on the distance 
from an estimate of the spectral type of the optical companion and quiescent 
optical observations and an upper limit from the flux of the X-ray bursts.
We study the X-ray variability on time scales from 0.008 to 4096~Hz and detect
high frequency quasi-periodic oscillations (QPOs), which were also reported by 
Homan et al.~(1998).  We derive model dependent estimates of the neutron star 
spin period and the magnetic field strength of the neutron star.  Finally, we 
study the relationship between the optical and X-ray flux during the decline 
from outburst to quiescence.  

\section{Observations}

The X-ray flux of XTE~J2123--058 during its 50 day outburst is shown in Figure~1.
The ASM light curve was produced from data provided by the ASM/RXTE teams at MIT 
and at the RXTE SOF and GOF at NASA's GSFC.  The peak ASM flux is about 
85 mCrab (1.5-12~keV).  Five pointed RXTE observations were made during the outburst, 
and the Proportional Counter Array (PCA) flux measurements for each observation are
shown in Figure~1.  Table~1 provides information about the pointed observations.
We performed spectral and timing analysis of the PCA data.

Figure~1 shows the V-band light curve for the nine nights we observed XTE~J2123--058
during the X-ray outburst.  Table~2 lists these observations along with an observation
made on September 20 after the X-ray outburst.  We used the 1.3 and 2.4 meter 
telescopes at the Michigan-Dartmouth-MIT Observatory (MDM), the 0.9 meter 
telescope at Kitt Peak National Observatory (KPNO) and the 0.9 meter telescope at 
Cerro Tololo Interamerican Observatory (CTIO).  The exposure times were between 
30 and 600~s.  The data were reduced using standard IRAF photometry reduction software.

\section{Optical Identification and Position}

Figure~2 shows two 6$^{\prime}$.5-by-6$^{\prime}$.5 V-band images of the XTE~J2123--058
field.  The top image is an average of three 5 minute exposures taken on 
June 30.  We identified the optical counterpart as a star in the 1$^{\prime}$ PCA 
error circle (\cite{takeshima98}) which had brightened to V = 16.9 from its quiescent 
level near the limiting magnitude, R = 21, of a digitized UK Schmidt Sky Survey plate 
(\cite{tomsick98a};~\cite{zurita98}).  The bottom image in Figure~2 shows a V-band image 
of the XTE~J2123--058 field taken on September 20 with the source probably in quiescence.  
We measured the optical position relative to six neighboring stars in the USNO A1.0 
astrometric catalog of the Palomar Observatory Sky Survey (\cite{monet96}).  The 
position of XTE~J2123--058 is R.A. = 21h 23m 14s.54, 
Decl. = --5$^{\circ}$~47$^{\prime}$~52.$^{\prime\prime}$9 (equinox 2000.0) 
with an uncertainty of 1$^{\prime\prime}$ (68\% confidence).  

\section{Optical Light Curve and Determination of the Orbital Ephemeris}

During outburst, the XTE~J2123--058 optical flux is highly modulated.  To determine 
the period of the modulation, we produced a Lomb periodogram (\cite{press92}).  
A highly significant peak occurs at a period of $5.9573\pm 0.0016$~hr.  Figure~3 
shows the V-band light curve from July 1 to July 16 folded on this period.  From 
July 1 to July 16, the X-ray and phase averaged optical flux levels were relatively 
constant, and the mean peak-to-peak amplitude of the modulation was about 0.75 
magnitudes.  From observations of standard stars made on June 30 (\cite{landolt92}), 
the mean V-band magnitude was 17.27 during this time.

It is likely that the optical modulation occurs because one side of the 
optical companion is X-ray heated.  In this case, high levels of modulation
are expected only if the binary inclination is relatively high.  Further
evidence for high binary inclination comes from the dip in the light curve
near an orbital phase ($\phi$) of 0.5, which is probably caused by a partial eclipse 
of the optical companion by the accretion disk.  Zurita et al.~(1998) observed
partial eclipses at $\phi = 0.5$ and also at $\phi = 0.0$ later in the outburst.
The photometric period we observe has been confirmed by spectroscopic measurements,
which indicates that it is the orbital period of the system (\cite{hynes98}).  

The light curve for each of six nights was fitted with a sine function
and the time of minimum light was found to within $\pm 0.00486$~d (68\% confidence).
For each night, the cycle count can be unambiguously determined, and 
we find that 172 orbits occurred between the first and last minima.
The ephemeris was found by fixing the period to the known value
and performing a $\chi^{2}$ minimization with the time of 
minimum light as the only free parameter.  The orbital ephemeris is 
HJD ($2451040.6581\pm 0.0034$) + ($0.248222\pm 0.000068$)E (90\% confidence errors).

\section{Binary Inclination}

The RXTE data were searched for modulation at the orbital period by folding the 
light curve for each observation on the orbital period.  We do not find evidence 
for modulation even though the RXTE observations provide complete phase 
coverage.  The absence of modulation puts an upper limit on the binary 
inclination ($i$).  Using the empirical relationship for systems with main 
sequence optical companions from Patterson (1984), the companion for a
5.9573~hr binary is expected to be a K8 star.  In the following, we assume that 
the spectral type of the optical companion is between M0 and K5.  We derive an 
upper limit on $i$ by assuming that the spectral type of the optical companion 
is M0 or earlier, and the mass of the neutron star is 1.4~M$_{\sun}$.  The no-eclipse 
condition for the neutron star is $i <$~cos$^{-1}(R_{2}/a)$, where $R_{2}$ is the 
radius of the optical companion's Roche lobe, which we derive from the relation 
given in Eggleton et al.~(1983), and $a$ is the binary separation.  The no-eclipse 
condition gives $i<73^{\circ}$.  A lower limit on $i$ can be derived if we interpret 
the dip in the optical light curve at $\phi = 0.5$ as being due to a partial eclipse 
of the optical companion by the accretion disk.  Assuming that a thin accretion disk
fills the neturon star's Roche lobe, the partial eclipse condition is 
$i >$~cos$^{-1}(R_{2}/(a - R_{1}))$, where $R_{1}$ is the radius of the neutron star's 
Roche lobe.  If the optical companion's spectral type is K5 or later, the inclination 
is greater than 55$^{\circ}$.

\section{X-Ray Bursts and Source Distance}

The times of the six X-ray bursts observed during the RXTE observations
are listed in Table 1, and the light curves for the three brightest bursts
are shown in Figure~4.  The bursts shown in Figures~4a and 4b were 
observed during observation 4, and the burst shown in Figure~4c was 
observed during observation 2.  The properties of the observation 4
bursts are considerably different from the four dimmer bursts.
Their rise times are 4~s compared to about 15~s for the dimmer bursts, 
and the observation 4 bursts are double-peaked, which may indicate that 
they are radius expansion bursts (\cite{lewin95}).  Since the peak burst 
luminosity is less than or equal to the Eddington luminosity ($L_{edd}$), 
a distance upper limit can be derived.  Using equation 4.10b of 
Lewin et al.~(1992) and assuming a neutron star mass of 1.4~M$_{\sun}$ 
and a neutron star atmosphere with a cosmic abundance of hydrogen gives 
$L_{edd} = 1.89\times 10^{38}$~erg~s$^{-1}$.  It should be noted that
$L_{edd}$ would be larger for a more massive neutron star or a neturon
star atmosphere with a lower hydrogen abundance.  We find the peak
bolometric flux by producing 1~s energy spectra near the peak of the
burst and fitting the spectra with a model consisting of the persistent 
emission plus a blackbody.  For the bursts shown in Figures 4a and 4b,
the peak bolometric fluxes are $6.8\times 10^{-9}$~erg~cm$^{-2}$~s$^{-1}$ 
and $7.1\times 10^{-9}$~erg~cm$^{-2}$~s$^{-1}$, respectively, giving a
distance upper limit of 15~kpc.  

Due to the high Galactic latitude of XTE~J2123--058, its distance is of 
considerable interest.  Here, we derive the distance from estimates of the 
spectral type of the optical companion and the quiescent magnitude of the 
source.  Observations made on August 26/27 gave R=$21.50\pm 0.06$ 
(\cite{zc98}), and we measured R=$21.60\pm 0.06$ on September 20.
There was no evidence for X-ray heating of the optical companion
during either observation.  Since X-ray heating was not observed and 
the source flux remained constant between August 26/27 and September 20,
the source was probably in quiescence.  Assuming that the companion is a 
K8 star (Patterson 1984) and that the quiescent emission is dominated by 
emission from the optical companion gives V = 22.52 and an absolute 
V-magnitude of 8.5, from which we derive a distance of 5.5~kpc.  
In deriving the quiescent V-band magnitude from the quiescent R-band
magnitude we use A$_{\rm V}$ = 0.30 and A$_{\rm R}$ = 0.24 to account 
for extinction (\cite{schlegel98}).  We note that the derived value for
V is consistent with our September 20 V-band measurement.  A range
of spectral types from M0 to K5 gives a distance range of 4.5 to 9.1~kpc.
We consider 4.5~kpc as a distance lower limit since this method underestimates 
the distance if light comes from the accretion disk in quiescence and conclude 
that the XTE~J2123--058 distance is between 4.5~kpc and 
15~kpc.  From the distance lower limit and the Galactic latitude, we find that the 
source must be at least 2.6~kpc from the Galactic plane.  This is unusual since 
the scale height for neutron star LMXB is only 1~kpc (\cite{vw95}).

Four optical bursts occurred during our observations.  On June 30, a burst of 
at least 0.15 V magnitudes occurred during a 60~s exposure.  We observed two 
bursts during the time from July 1 and July 16, which are shown in Figure~3.
One of these bursts produced increases of 0.38 and 0.10 magnitudes for two 
consecutive 30~s exposures and the other produced a 0.21 magnitude increase 
during a 60~s exposure.  The largest optical burst was observed during the first 
exposure taken on August 6.  As shown in Figure~9, the 120~s burst exposure was 
0.47 magnitudes brighter than the following exposures.  Although we speculate that 
these bursts are reprocessed X-ray bursts, there were no simultaneous X-ray 
observations during any of the optical bursts.  To see if the data are consistent 
with this hypothesis, we compared the energy released during X-ray bursts to 
that during the optical bursts.  For a burst where the V magnitude increases from 
16.9 to 16.5 and lasts 30~s, the burst fluence in the V-band is 
$8.5\times 10^{-12}$~erg~cm$^{-2}$.  The X-ray fluence for an average 
XTE~J2123--058 burst is $2.3\times 10^{-8}$~erg~cm$^{-2}$ giving an optical to 
X-ray burst fluence ratio of $4\times 10^{-4}$.  For the persistent emission, we 
find L$_{opt}$/L$_{x}$ ratios between $3\times 10^{-4}$ and $1.5\times 10^{-3}$.  
The similarity of the persistent and the burst L$_{opt}$/L$_{x}$ ratios indicates 
that the optical bursts are likely to be reprocessed X-ray bursts.

\section{High Frequency Quasi-Periodic Oscillations}

To search for QPOs, we made 0.5-4096~Hz power spectra from the RXTE data.
We produced 2~s Leahy normalized power spectra (\cite{leahy83}) using
the available data in the 2-20~keV energy band.  For each RXTE orbit, 
the 2~s spectra were combined and we looked for QPOs in the combined spectra.
For observation 1, it is not possible to carry out this analysis because
there are no high time resolution data.  We only detect QPOs during observation 
3, which confirms the result of Homan et al.~(1998).  A QPO at $849.4\pm 2.7$~Hz 
is detected during the first orbit of this observation and a QPO at 
$1141.1\pm 6.5$~Hz is detected during the second orbit.  The errors given in this 
section are 68\% confidence.  QPOs are not detected for the other orbits.  Fitting 
the power spectrum with a model consisting of a constant plus a Lorentzian, we 
find that the 849~Hz QPO has a width (FWHM) of $18.0\pm 4.9$~Hz and a fractional 
rms amplitude of $7.4\pm 0.7$~\%, and the 1141~Hz QPO has a width of 
$30.9\pm 13.4$~Hz and a fractional rms amplitude of $7.8\pm 1.1$~\%.  F-tests 
indicate that both QPOs are detected at greater than 99\% confidence.
QPOs are observed during X-ray bursts from several systems (\cite{s98}).  
We searched for, but do not detect, QPOs in the brightest two X-ray bursts.

With the detection of kHz QPOs in XTE~J2123--058, this source joins a group of 18
other neutron star LMXBs with high frequency QPOs (\cite{v98}).  For several of these 
systems, two high frequency QPOs are observed simultaneously.  According to the beat 
frequency model, the QPO with the higher frequency corresponds to the Keplerian frequency 
at the inner edge of the accretion disk and the QPO with the lower frequency corresponds 
to the beat frequency between the Keplerian frequency and the spin frequency of the 
neutron star.  In this picture, the difference between the two QPO frequencies is the 
spin frequency of the neutron star (\cite{alpar82}).  This interpretation is supported 
by the fact that, in several sources, the difference between the two QPO frequencies is 
approximately constant even though the QPO frequencies change (see e.g. \cite{ford97}).  
In addition to the detection of the 849~Hz QPO during the first RXTE orbit of
observation 3, there is marginal evidence for a QPO at $1104\pm 15$~Hz.  The feature
is significant at the 86\% confidence level.  The detection is significant at the 
92\% confidence level if we use only the last 2240~s of the first orbit data rather 
than the entire 3344~s.  The detection can be further improved to 99\% confidence
if the 4.6-20~keV energy band is used rather than the 2-20~keV energy band.  This
is consistent with the result of Homan et al.~(1999) that the QPO strength increases
with photon energy.  Figure~5 shows the 4.6-20~keV power spectrum for the last 2240~s 
of the first orbit fitted with a model consisting of a constant and two Lorentzians.  
Both QPOs are detected at greater than 99\% confidence.  The QPO frequencies are 
$847.1\pm 5.5$~Hz and $1102\pm 13$~Hz with fractional rms amplitudes of $10.3\pm 1.4$~\% 
and $11.8\pm 1.6$~\%, respectively.  According to the beat frequency model, the observed 
frequency difference of $255\pm 14$~Hz implies a neutron star spin period of 
$3.92\pm 0.22$~ms.  However, we note that the validity of the simple beat 
frequency model is in question because in some sources the difference between the two 
QPOs is not constant (\cite{v97};~\cite{mendez98}).  In fact, Psaltis et al.~(1998) 
suggest that the QPO separation is not constant for any source with two kHz QPOs.

\section{X-Ray Energy Spectra}

For the first four RXTE observations, we produced 2.5--20~keV PCA energy spectra 
using the processing methods described in Tomsick et al.~(1998c).  For observation 5,
the data were processed in the same way except that the background spectrum was estimated
using a model for faint sources (\cite{stark98}).  Only the photons from the top anode
layers were used.  As described in Tomsick et al.~(1998c), Crab spectra were made 
to test the response matrices.  Of the five Proportional Counter Units (PCUs),
the response matrix is best for PCUs 1 and 4.  Since PCU 4 was off during
observation 1 and part of observation 2, the spectral fits were carried out using only
PCU 1 in order to avoid instrumental differences between observations.  Also, fluxes or 
spectral component normalizations were reduced by a factor of 1.18 so that the PCA 
flux scale is in agreement with previous instruments (\cite{tomsick98c}).  Energy 
spectra for the persistent emission were produced using data with 16~s time 
resolution.  Time bins which overlap with X-ray bursts as specified in Table~1
were excluded.  When the first two observations were made the source had not been 
optically identified and the source was about 0.2 degrees from the center of the 
field of view.  The June 5, 1996 collimator response was used to correct for 
this pointing offset.

The spectra were fitted with a power-law model, a Comptonization model (\cite{st80})
and a model consisting of a blackbody plus a power-law.  For each model, the fit
was calculated with the column density free and also with the column density
fixed to the Galactic value ($5.73\times 10^{20}$~cm$^{-2}$).  
A power-law alone provides a very poor fit to the first four spectra, but for the
fifth spectrum, a power-law with a photon index ($\Gamma$) of 1.98 provides
an acceptable fit ($\chi^{2}/\nu$ = 36/45 with the column density free and 
$\chi^{2}/\nu$ = 37/46 with the column density fixed).  The fits to the first
four spectra are improved by using a Comptonization model or a model consisting of 
a blackbody and a power-law.  The fit parameters for these two models are given in 
Table~3.  For the first four spectra, Comptonization models with temperatures ($kT$) between 
2.5 and 3.0~keV and optical depths ($\tau$) between 11 and 14 provide reasonably good 
fits to the data.  For observation 5, the Comptonization model does not 
provide a better fit than the power-law model, and the Comptonization parameters are 
very poorly constrained.  The column density is consistent with the Galactic
value for observation 5 if a power-law or a Comptonization model is used.
For observations 2, 3 and 4, F-tests indicate that the spectra require a 
column density different from the Galactic value at greater than 90\% confidence
if a Comptonization model is used.

The blackbody plus power-law fits give similar results for the first four spectra, but
the fitted parameters are significantly different for observation 5.
For the first four spectra with the column density free, the blackbody temperatures are 
between 1.86 and 1.94~keV, the values for the photon index, $\Gamma$, are between 2.5 
and 3.1, and the column densities are between 1.3 and 2.6~$\times 10^{22}$~cm$^{-2}$.
For observation 5, the blackbody temperature is 
0.65~keV, $\Gamma = 1.85$, and the column density is consistent with the Galactic 
column density.  For observations 2, 3 and 4, F-tests indicate that the spectra
require a column density different from the Galactic value at greater than 
99\% confidence.

Near the end of the X-ray outburst from Aql~X--1 in early 1997, Zhang et al.~(1998) 
observed a sudden change in the hardness of the X-ray spectrum.  They interpret 
the change as the onset of the ``propeller'' mechanism (\cite{ill75}).  In this
picture, as the mass accretion rate onto the neutron star drops, the size of the
neutron star magnetosphere increases until it produces a centrifugal barrier
for the accreting material.  Zhang et al.~(1998) fit the Aql~X--1 energy spectra 
with a model consisting of a blackbody, a power-law and an iron line and find two 
main changes after the transition:  the blackbody temperature decreases and the 
power-law index becomes harder.  For XTE~J2123--058, the power-law plus blackbody 
fit parameters for observations 4 and 5 show a similar change.  The 
energy spectra for observations 4 and 5 fitted with a blackbody plus 
power-law model with the column density free are shown in Figure~6.  Between 
observations 4 and 5 the blackbody temperature changed from $1.91\pm 0.02$~keV 
to $0.67^{+0.06}_{-0.16}$~keV (68\% confidence errors) and the power-law index 
changed from $3.06\pm 0.05$ to $1.85\pm 0.07$.  Assuming that the change is due 
to the onset of the propeller mechanism, we can estimate the strength of the 
neutron star magnetic field using equation 2 of Zhang et al.~(1998).  For this 
estimate, we assume 1.4~M$_{\sun}$ and $10^{6}$~cm for the mass and radius of the 
neutron star.  Assuming that the neutron star spin period is 3.92~ms and that 
the critical X-ray luminosity for the transition is between the observation 4 and 
5 levels, the neutron star magnetic field strength is between 2 and 
8~$\times 10^{8}~d_{10}$~G, where $d_{10}$ is the source distance in units of 10~kpc.

\section{Correlated Spectral and Timing Behavior}

For observations 2 to 5, we produced 0.008-128~Hz power spectra to study the
timing properties of the system.  An rms normalized power spectrum was made for
each 128~s interval using non-burst data in the 2-20~keV energy band.  To convert 
from the Leahy normalization to rms, we determined the Poisson noise level using the 
method described in Zhang et al.~(1995) with a deadtime of 10 microseconds per event.  
The background count rates were estimated as described in the previous section.
For each of the four observations, the 128~s power spectra were combined and then 
fitted with a power-law model.  For observations 2 and 5, a power-law provides an
acceptable fit to the data; however, for observations 3 and 4, we detect excess
power above 5~Hz.  For observations 3 and 4, the power spectra were
fitted with a model consisting of a power-law plus a cutoff power-law.  This
model has the form
\begin{equation}
P(\nu) = A_{1}~\nu^{\alpha_{1}} + A_{2}~\nu^{\alpha_{2}}~e^{-\frac{\nu}{\nu_{c}}}~,
\end{equation}
where $A_{1}$, $\alpha_{1}$, $A_{2}$, $\alpha_{2}$ and $\nu_{c}$ are free parameters.
For all four observations, the fitted parameters are given in Table~4.  Instead of
reporting the values for $A_{1}$ and $A_{2}$, fractional rms amplitudes are given
for the two components.

During the outburst there were two significant changes in the timing properties of 
the source.  The first change occurred between observations 2 and 3 when excess noise 
above 5~Hz appeared in the spectrum.  For observations 3 and 4, the fractional rms 
amplitudes (0.01-100~Hz) for the cutoff power-law component are $5.5\pm 0.4$~\% 
and $4.4\pm 0.4$~\%, respectively (68\% confidence errors).  For observation 2, 
the 90\% confidence upper limit on the fractional rms amplitude of a similar component 
is 1\%.  The second change occurred between observations 4 and 5.  For observations 
2, 3 and 4 the fractional rms amplitudes are between 2.2\% and 2.8\%, while for 
observation 5, the fractional rms amplitude is $21.2\pm 3.5$~\%.  
The power spectra for observation 2, observations 3 and 4 combined and 
observation 5 are shown in Figures~7a, 7b and 7c, respectively.  Although a 
power-law provides an acceptable fit for observation 5 ($\chi^{2}/\nu$ = 10.7/12), 
the observation 5 power spectrum was re-fitted with the function given in equation~1.
The fitted cutoff power-law parameters are $\alpha = 0.5$ and $\nu_{c} = 7$~Hz.
The 90\% confidence upper limit on the fractional rms amplitude of the cutoff 
power-law component is 21\%.

We made a color-color diagram (CD) to see how the changes in the XTE~J2123--058
timing properties are related to its spectral properties.  Figure~8 shows the CD, 
where the ratio of the 3.5-6.4~keV to 2.1-3.5~keV count rates is plotted on the
horizontal axis (soft color), and the ratio of the 9.8-20~keV to 6.4-9.8~keV
count rates is plotted on the vertical axis (hard color).  The source occupied 
different regions of the CD during the outburst.  The CD suggests that XTE~J2123--058 
should be classified as an atoll source (\cite{v95}), and that the source was in the 
upper banana state (UB) during observations 1 and 2, the lower banana state (LB) during 
observations 3 and 4 and the island state during observation 5 (\cite{v95}).  This 
interpretation is supported by the timing properties of the source during observations
2 to 5.  A change from the UB state to the LB state is typically accompanied by 
an increase in the high frequency noise component (which we characterize using a cutoff 
power-law).  Also, the increase in the rms amplitude we observe between observations 
4 and 5 defines the transition from the banana state to the island state.  That 
XTE~J2123--058 is an atoll source is not surprising since X-ray bursts are commonly 
seen in atoll sources.

The timing analysis presented here causes us to question whether the change
observed during the outburst decay is related to the onset of the propeller mechanism.  
Based on its CD and the timing properties during the decay, the behavior of 
XTE~J2123--058 appears to be typical for atoll sources as the mass accretion rate 
onto the neutron star decreases.  It is not clear if the spectral and timing
changes are due solely to the dynamics of the disk or due to the interaction 
between the disk and the magnetosphere.  The latter possibility may suggest 
that the onset of the propeller mechanism causes the change from the banana to 
the island state in atoll sources.  A argument against this is that the X-ray 
timing properties for black hole systems, where the compact object does not have 
an intrinsic magnetosphere, are similar to atoll source timing properties (\cite{v95}).

\section{Relationship between Optical and X-Ray Flux}

We observed XTE~J2123--058 in the optical at X-ray flux levels from quiescence to 
$1\times 10^{-9}$~erg~cm$^{-2}$~s$^{-1}$ (1.5-12~keV).  Our optical observations
provide sufficient phase coverage to study the shape of the optical light curve
at three X-ray flux levels.  Table~5 gives the mean X-ray flux levels for the 
period of time between July 1 and July 16, on August 6 and on August 15.
The folded optical light curve for July 1-16 is shown in Figure~3, and Figure~9
shows the V-band light curves for August 6 and August 15.  For the August 
observations, we obtained good, but not complete, phase coverage.  For July 1-16
and August 6, the X-ray fluxes come from RXTE/ASM measurements.  Since the
fifth pointed RXTE observation began immediately after our August 15 optical
observation, the PCA flux measurement was used.

For each time interval, we fitted the optical light curve with a sine function 
to determine the V-band magnitude at maximum light ($\phi = 0.5$) and
minimum light ($\phi = 0.0$).  The modulation amplitude (V(0.5)--V(0.0)) increases 
as the X-ray flux decreases.  This trend was also observed in the R-band by
Zurita et al.~(1998).  To understand this trend, we assume that the optical flux 
can be separated into a modulated component coming from the optical companion and 
an unmodulated component coming from the accretion disk.  To separate the
components, the V-band magnitudes given in Table~5 were converted into flux 
densities (i.e. erg~cm$^{-2}$~s$^{-1}$~Hz$^{-1}$ at $5.455\times 10^{14}$~Hz) 
using the relationship given in Table IV of Bessell et al.~(1979).  A flux density 
was also determined from the estimated quiescent V-band magnitude of 22.52.  To 
correct for extinction, A$_{V}$=0.3 was used.  The peak contribution to the flux 
density from the optical companion is given by 
$F_{\rm V,OC\it} = F_{\rm V\it}(0.5)-F_{\rm V\it}(0.0)$, where $F_{\rm V\it}(\phi)$ is 
the V-band flux density at the orbital phase $\phi$.  The contribution to the flux 
density from the disk is given by $F_{\rm V,D\it} = F_{\rm V\it}(0.0)-F_{\rm V,Q\it}$, 
where $F_{\rm V,Q\it}$ is the quiescent flux density.  The values for
$F_{\rm V,OC\it}$ and $F_{\rm V,D\it}$ are given in Table~5.  From July 1-16 to 
August 15, $F_{\rm V,D\it}$ decreased by a factor of 10, while $F_{\rm V,OC\it}$ 
only decreased by a factor of 3.  In the following, we assume that the optical light 
comes from X-ray reprocessing in the accretion disk and the optical companion.
First, we calculate the expected change in $F_{\rm V,OC\it}$ as the X-ray flux 
drops from its July 1-16 level to its August 15 level, and find that it
is consistent with the observed factor of 3.  Then, a simplified model of
the system is used to understand the physical changes in the system during
the outburst decay.

We calculate the X-ray flux from the heated side of the optical companion
by determining the temperature, $T$, at each point on the star using equation 1.8
of van Paradijs~(1991).  Assuming that each surface element emits as a blackbody at 
the local temperature, the integral over the surface of the star is
\begin{equation}
F_{\rm V,OC\it} = 2\pi~\int_{0}^{\theta_{c}} \theta~B_{\rm V\it}[T(\theta)]d\theta~,
\end{equation}
where $\theta_{c}$ is the ratio of the neutron star radius to the source distance.
For a binary at 10~kpc consisting of a 1.4~M$_{\sun}$ neutron star and a K8 star 
with an X-ray albedo of 0.4 (\cite{london81}), the July 1-16, August 6, and 
August 15 X-ray fluxes imply $F_{\rm V,OC\it}$ values of 
$7.7\times 10^{-27}$~erg~cm$^{-2}$~s$^{-1}$~Hz$^{-1}$, 
$5.6\times 10^{-27}$~erg~cm$^{-2}$~s$^{-1}$~Hz$^{-1}$, and
$2.3\times 10^{-27}$~erg~cm$^{-2}$~s$^{-1}$~Hz$^{-1}$, respectively.  
Although these values are a factor of about 2 higher than the observed
$F_{\rm V,OC\it}$ values, the ratio between the July 1-16 value and the 
August 15 value is about 3, consistent with observations.  There are
several reasons why our calculated $F_{\rm V,OC\it}$ values are higher than 
the observed values.  First, to simplify the calculation, we assume
a binary inclination ($i$) of 90$^{\circ}$.  Assuming that $i$ is not too much
less than the upper limit of 73$^{\circ}$, this correction changes the calculated 
value by a factor of sin~$i$.  As stated above, the dip in the light curve near 
orbital phase 0.5 probably indicates that the accretion disk blocks part of 
the optical companion at maximum light, but our calculation assumes that we 
observe the entire optical companion unobstructed.  Finally, if the
accretion disk is flared at its edge (\cite{wh82}), the disk may 
partially shield the optical companion from the X-ray source, reducing the
level of X-ray heating.

The values of $F_{\rm V,D\it}$ can be used to understand the
evolution of the accretion disk during the decay of the outburst.
We assume a spherical X-ray source of radius $R$ and a thin, azimuthally 
symmetric accretion disk extending radially from $R$ to $R_{out}$ and 
inclined at an angle $i$ with respect to the line of sight of the observer.  
Further, we assume that the V-band emission
is dominated by blackbody emission from the accretion disk at a
temperature $T$ and that the temperature is only a function of the
radial coordinate, $r$.  Based on UV measurements of other LMXBs,
we expect the accretion disk temperature to be high enough so that
the V-band lies in the Rayleigh-Jeans portion of the blackbody spectrum
(\cite{vm95}).  At a frequency $\nu$, the flux density is
\begin{equation}
F_{\nu,\rm D\it} = \left(\frac{4\pi~\nu^{2}~k~\rm~cos\it~i}{c^{2}~d^{2}}\right)~
\int_{R}^{R_{out}} T(r)~r~dr~,
\end{equation}
where $c$ is the speed of light, $k$ is Boltzmann's constant and $d$ is
the distance to the source.  If the V-band emission comes from
X-ray reprocessing, the temperature can be written as
\begin{equation}
T(r) = \left(\frac{F_{x}~R~d^{2}~(1-\eta)}{2~\sigma~r^{3}}\right)^{\frac{1}{4}}~,
\end{equation}
where $\eta$ is the albedo of the disk.  Equation~4 assumes that $R$ is much 
less than $r$.  Although we integrate over values of $R$ to $R_{out}$, this 
approximation is valid because most of the flux comes from close to $R_{out}$.  
Assuming that $R$ is much less than $R_{out}$, the V-band flux density from the disk is
\begin{equation}
F_{\rm V,D\it} = 4.062\times 10^{-28}~F_{x,-10}^{\frac{\rm 1}{\rm 4}}~
\rm~cos\it~i~d_{\rm 10}^{-\frac{\rm 3}{\rm 2}}~(\rm 1-\eta)_{\rm 0.6}^{\frac{\rm 1}{\rm 4}}~
\it R_{\rm 7}^{\frac{\rm 1}{\rm 4}} \it R_{out,\rm 10}^{\frac{\rm 5}{\rm 4}}~\rm erg~cm^{-2}~s^{-1}~Hz^{-1}~,
\end{equation}
where $F_{x,-10}$ is the X-ray flux in units of 
$10^{-10}$~erg~cm$^{-2}$~s$^{-1}$, $R_{7}$ is the radius of the X-ray 
source in units of $10^{7}$~cm, $R_{out,10}$ is the outer radius of the disk 
in units of $10^{10}$~cm and $d_{10}$ is the distance to the source in units 
of 10~kpc.  Equation~5 can be used to understand how the geometry of the 
X-ray source and the disk change during the X-ray decline.  
Using the values of $F_{\rm V,D\it}$ and $F_{x,-10}$ from Table~5, 
cos~$i$~$d_{10}^{-\frac{3}{2}}$~$(1-\eta)_{0.6}^{\frac{\rm 1}{\rm 4}}$~
$R_{7}^{\frac{1}{4}} R_{out,10}^{\frac{5}{4}}$ is equal to 5.8, 2.7, and 
1.2 for July 1-16, August 6 and August 15, respectively.  Only $R$ and
$R_{out}$ are likely to change during the outburst decay.  $R_{out}$ 
decreases by a factor of 3.6 between July 1-16 and August 15 if $R$ is 
constant during the X-ray decay.  If instead we assume that the X-ray
flux is proportional to the surface area of the spherical X-ray emitting
region (i.e. $F_{x}$ is proportional to $R^{2}$), then $R_{out}$ still 
decreases by a factor of 2.8 between July 1-16 and August 15.
We conclude that the outer radius of the accretion disk decreases as the
X-ray flux decreases.

\section{Summary and Conclusions}

We optically identified the neutron star X-ray transient XTE~J2123--058,
measured its orbital period and constrained the binary inclination.  The 
orbital period is $5.9573\pm 0.0016$~hr and the binary inclination is 
between 55$^{\circ}$ and 73$^{\circ}$.  The properties of the observed 
optical bursts are consistent with being reprocessed X-ray bursts.
The distance to the source is between 4.5~kpc and 15~kpc, indicating that 
the source is at least 2.6~kpc from the Galactic plane, which is 
considerably greater than the 1~kpc scale height for neutron star LMXBs.  
To determine the distance lower limit, we have assumed that all the 
quiescent optical light comes from a main sequence optical companion.  

High frequency QPOs at $847.1\pm 5.5$~Hz and $1102\pm 13$~Hz are simultaneously 
detected.  According to the beat frequency model, the difference in frequency 
between the two QPOs indicates a neutron star spin period of $3.92\pm 0.22$~ms.  
During the decay of the outburst, a change in the energy spectrum is observed, 
which is similar to a change observed in Aql~X-1.  We speculate that this is 
due to the onset of the propeller mechanism, indicating that the neutron star
magnetic field strength is in the range from 2 to 8~$\times 10^{8}~d_{10}$~G.
However, the changes in the timing and spectral properties observed during
the decay are typical of atoll sources, which may indicate that the changes
are due solely to the dynamics of the accretion disk.

As the X-ray flux decays, we infer that the V-band flux density from the accretion
disk drops by a factor of 10, while the V-band flux density from the optical 
companion only changes by a factor of 3.  Assuming the optical flux comes from 
X-ray reprocessing in the disk and the optical companion, our calculations show
that the V-band flux density from the optical companion is expected to change
by a factor of 3 for the observed change in X-ray flux and that the factor of
10 change in the V-band flux density from the accretion disk can be explained
if the size of the accretion disk decreases significantly during the decay of
the outburst.

Although most of the properties of XTE~J2123--058 are typical for neutron
star LMXBs, its distance from the Galactic plane is unusual.  In the future,
it is important to improve constraints on its distance.  This can be 
accomplished from a spectroscopic determination of the spectral type of 
the optical companion.  Also, the low Galactic column density for this 
system provides a good opportunity to study the soft X-ray properties of
a neutron star transient.  This will be especially useful for X-ray 
observations in quiescence (e.g. \cite{rutledge98}).

\acknowledgements

The authors would like to thank D. Hurley-Keller for observing XTE~J2123--058, 
K. Leighly for assisting with our observations and J. Orosz for useful
discussions.  P.K. and J.A.T. acknowledge support from NASA grants NAG5-3595, 
NAG5-3799, NAG5-4416, and NAG5-7405.

\clearpage

%FIGURES

\begin{figure} \figurenum{1} \epsscale{0.9} \plotone{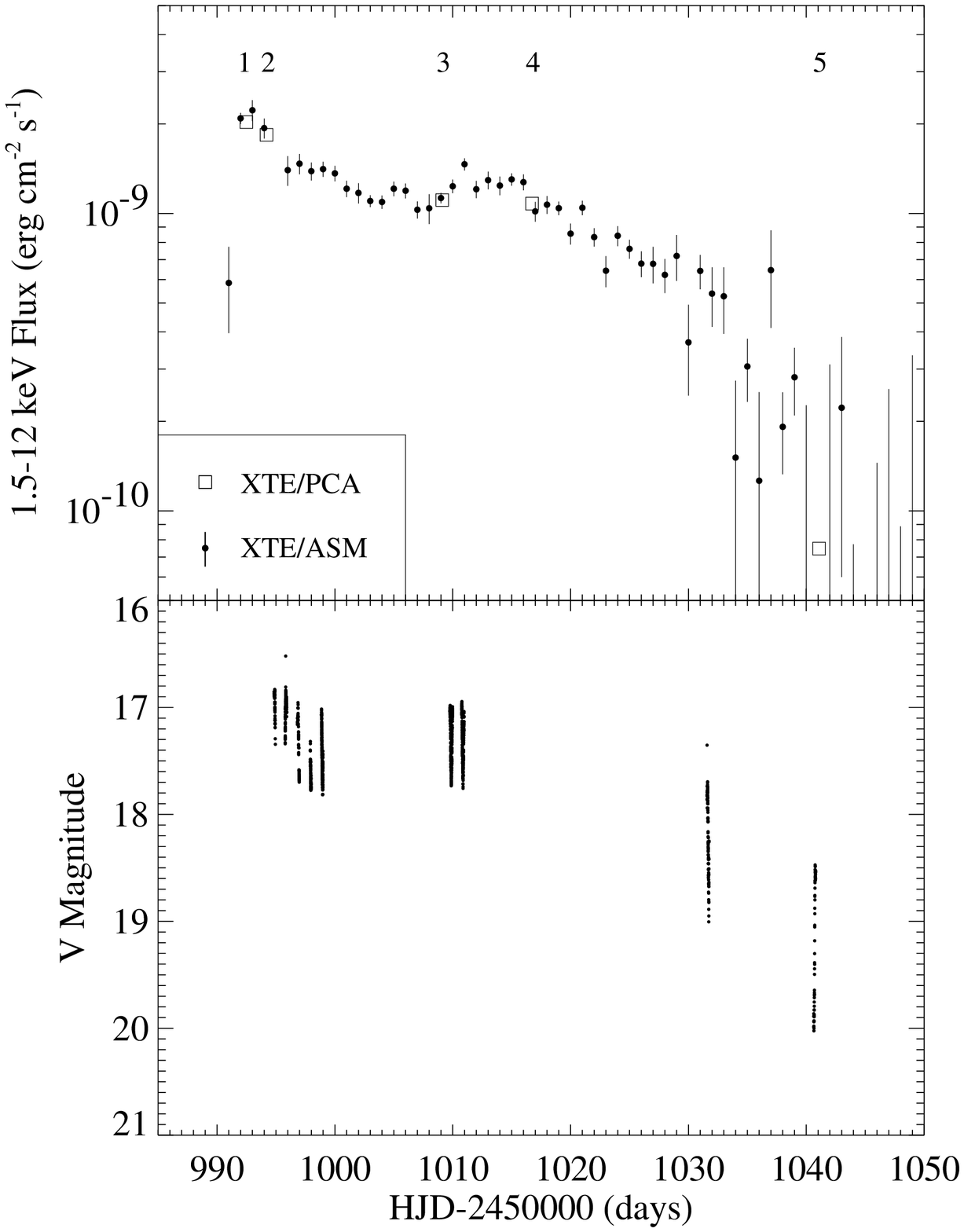}
\caption{X-ray (top) and V-band (bottom) light curves for the XTE~J2123--058
outburst.  The points with error bars are 1.5-12~keV ASM daily flux measurements, and 
the squares mark 1.5-12~keV flux measured by the PCA during each of the five pointed 
RXTE observations.\label{fig1}}
\end{figure}

\begin{figure} \figurenum{2} \epsscale{0.8} \plotone{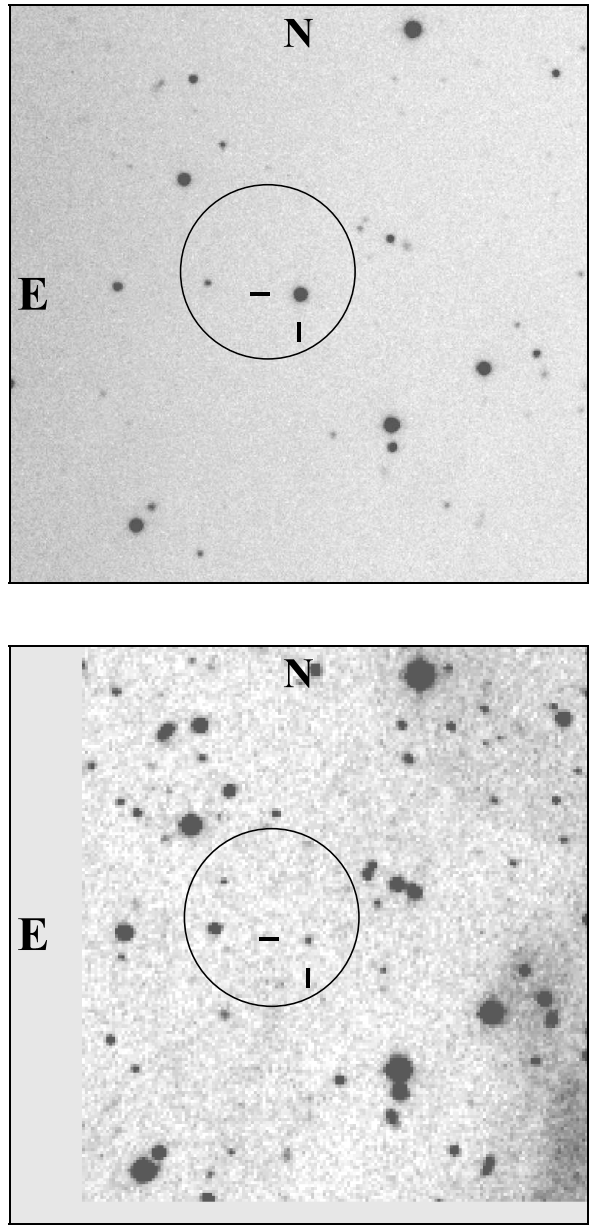}
\vspace{2.0cm}
\caption{Finding charts for XTE~J2123--058 in outburst (V=16.9) and 
quiescence.  The optical position is R.A. = 21h 23m 14s.54, 
Decl. = --5$^{\circ}$~47$^{\prime}$~52.$^{\prime\prime}$9 (equinox 2000.0).  The
charts are 6$^{\prime}$.5-by-6$^{\prime}$.5 V-band images and the 1$^{\prime}$ PCA 
error circle is shown.\label{fig2}}
\end{figure}

\begin{figure} \figurenum{3} \epsscale{1.0} \plotone{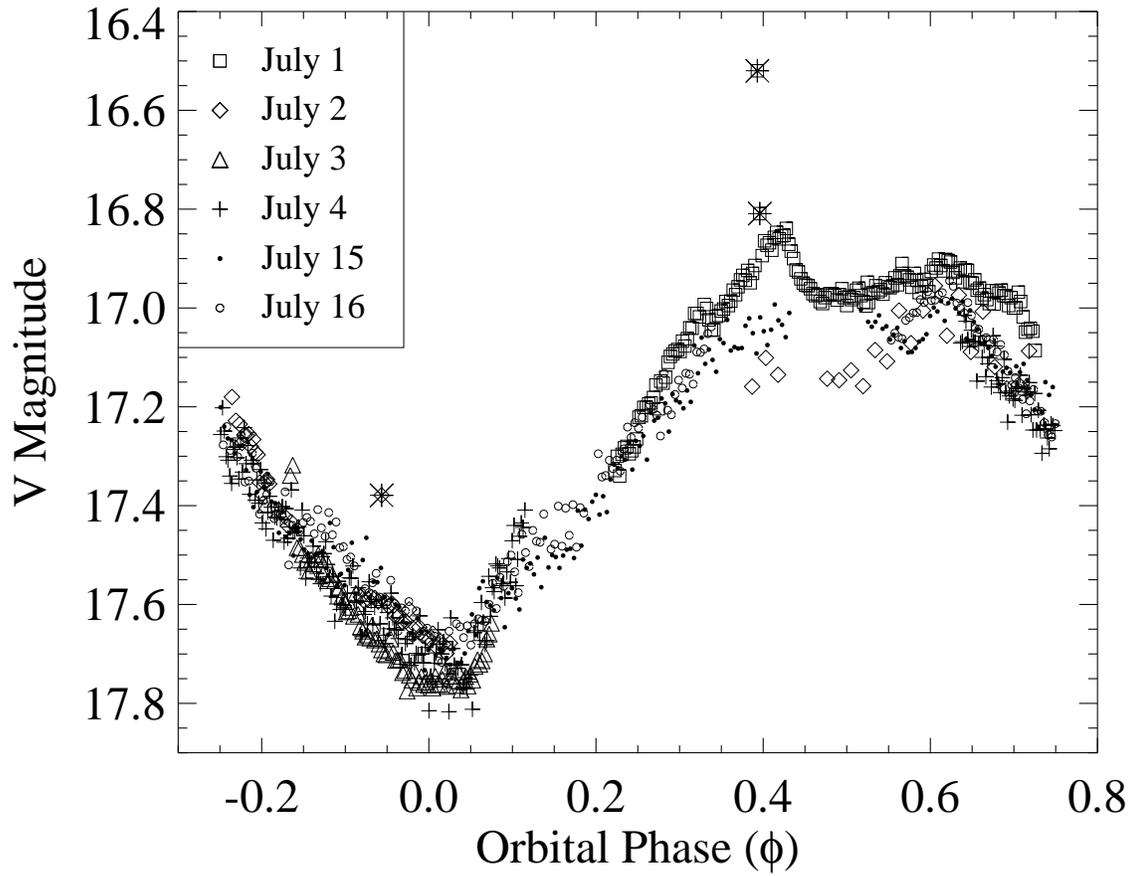}
\caption{The V-band light curve from July 1 to July 16 folded on the 5.9573~hr 
orbital period.  Each symbol corresponds to a different night as indicated in the legend.
Optical bursts are marked with asterisks.\label{fig3}}
\end{figure}

\begin{figure} \figurenum{4} \epsscale{0.9} \plotone{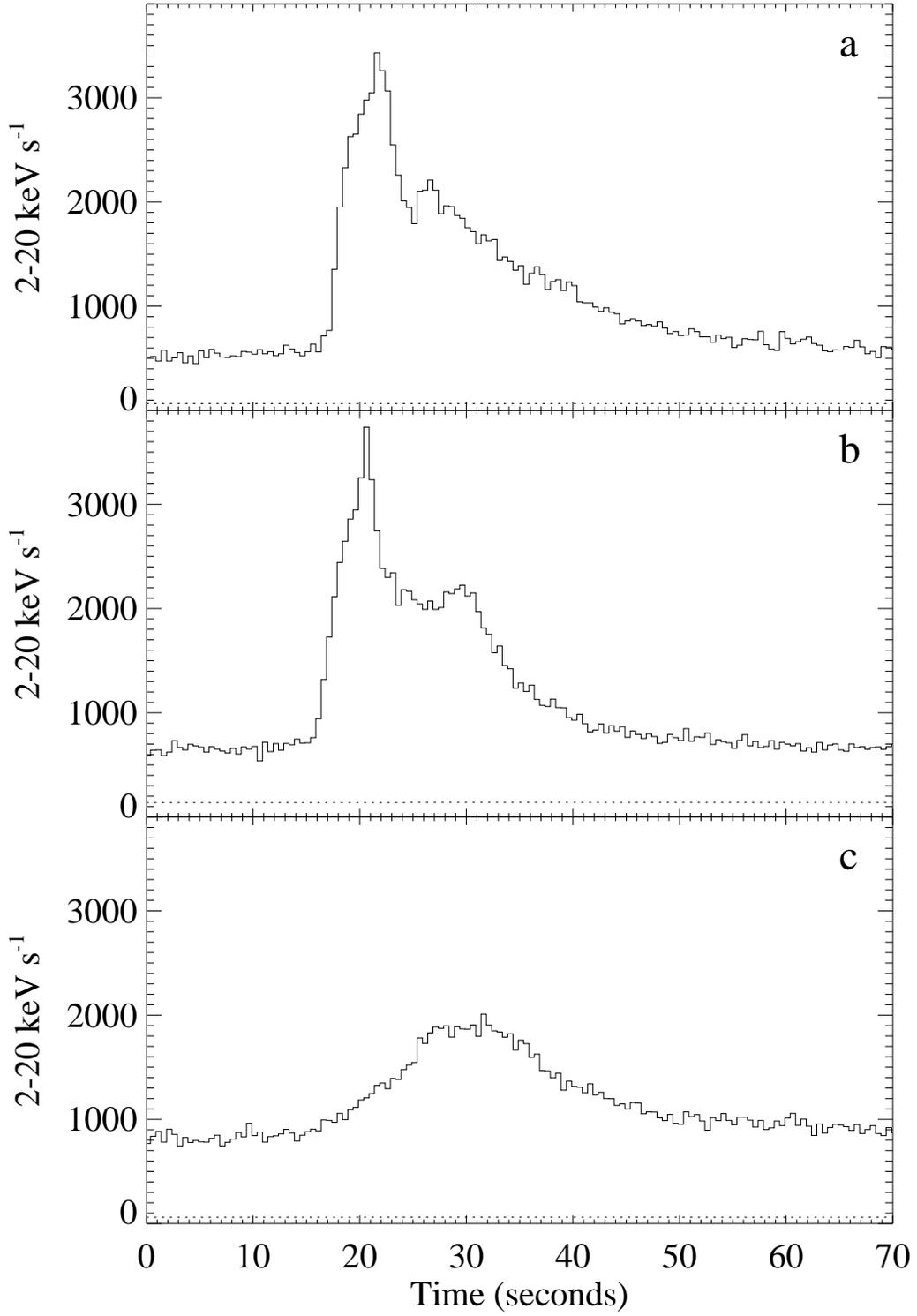}
\vspace{1.5cm}
\caption{(a) and (b) PCA light curves for the two brightest bursts; and 
(c) The PCA light curve for the burst observed during observation 2.  In each plot, 
the dotted line indicates the background level.\label{fig4}}
\end{figure}

\begin{figure} \figurenum{5} \epsscale{1.0} \plotone{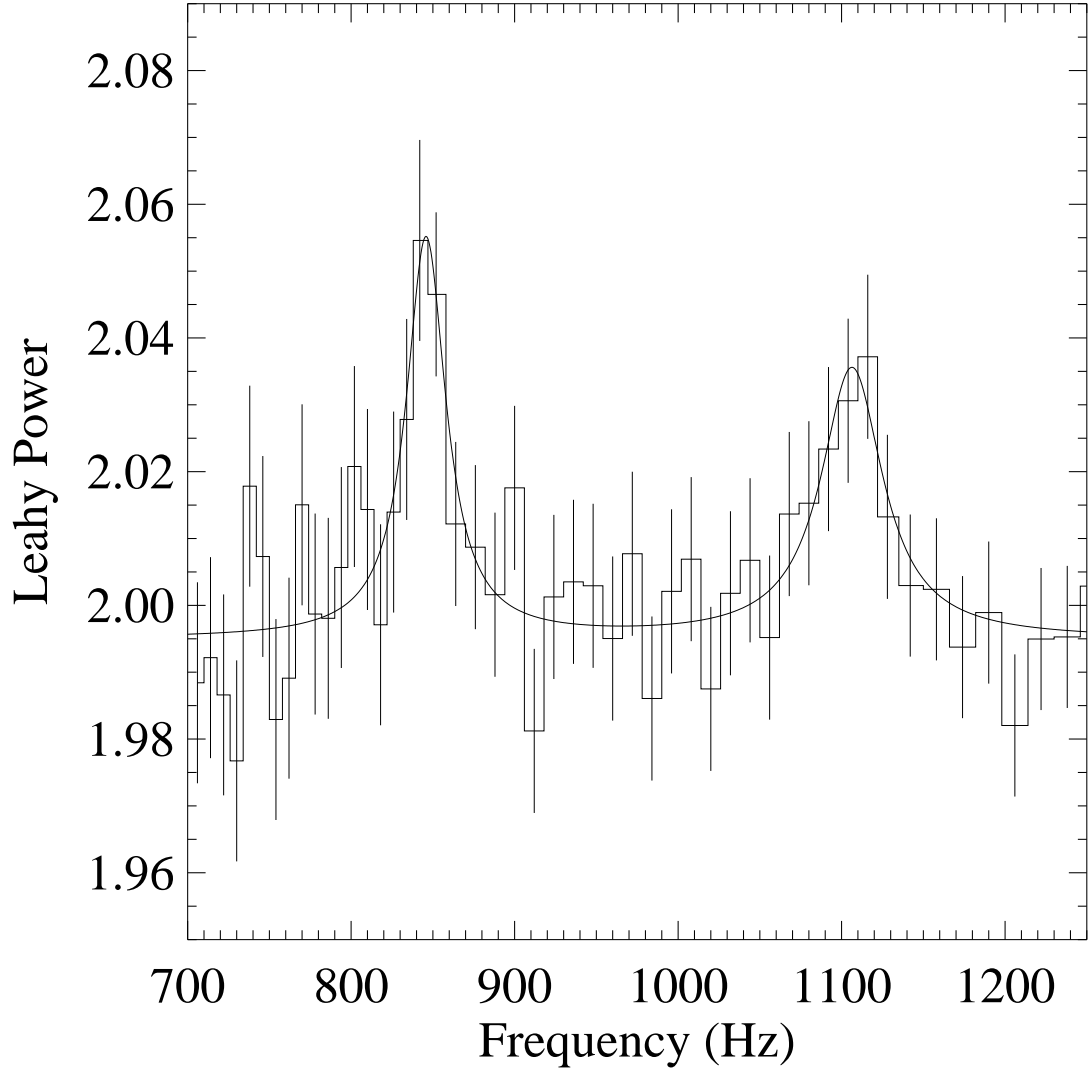}
\caption{Power spectrum for the last 2240~s of the first orbit of
observation 3.  The 4.6-20~keV energy band is used and QPOs are detected
at $847.1\pm 5.5$~Hz and $1102\pm 13$~Hz.\label{fig5}}
\end{figure}

\begin{figure} \figurenum{6} \epsscale{0.9} \plotone{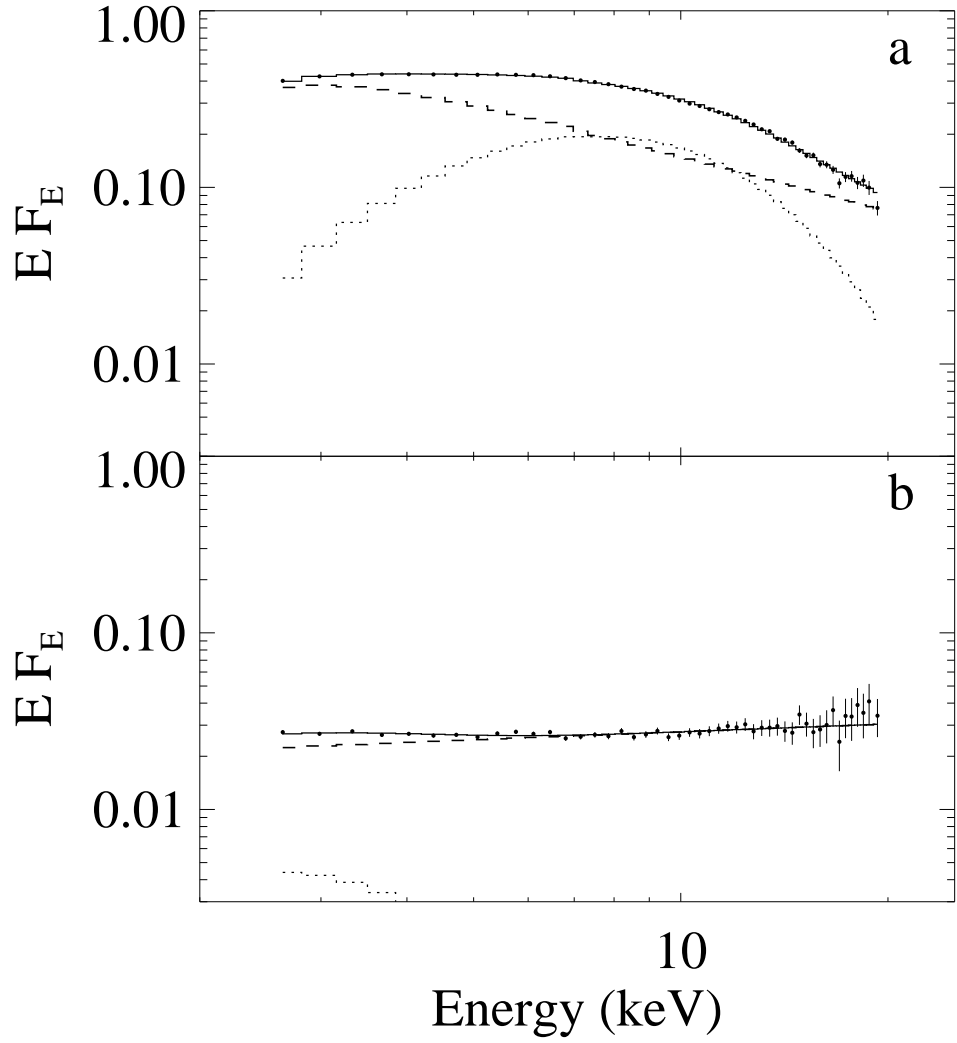}
\vspace{1.5cm}
\caption{Energy spectra for observations 4 and 5 fitted with a blackbody
(dotted line) plus power-law (dashed line) model.  The solid line is the sum of the
two components.\label{fig6}}
\end{figure}

\begin{figure} \figurenum{7} \epsscale{0.9} \plotone{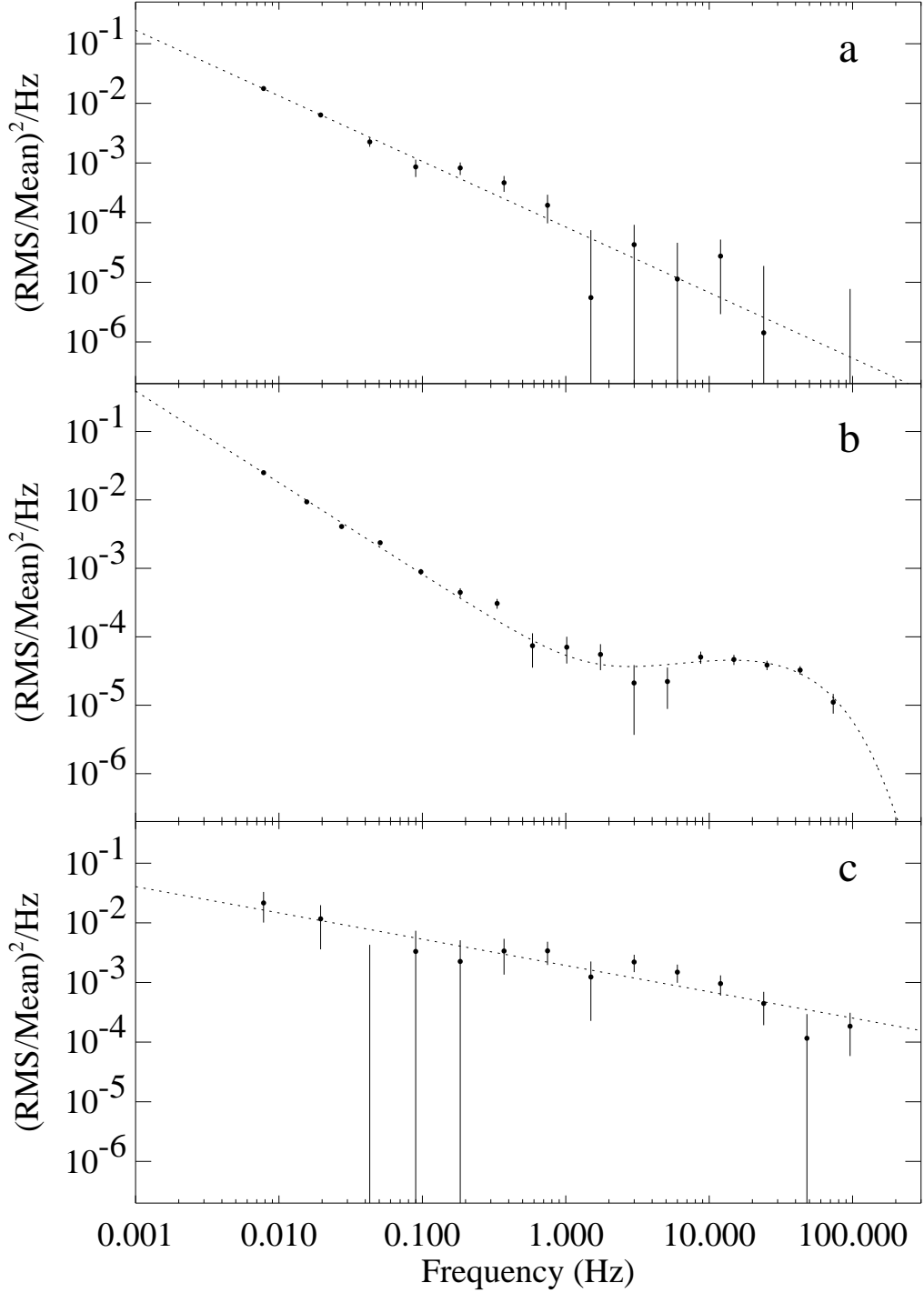}
\vspace{1.5cm}
\caption{RMS normalized 2-20~keV power spectra for (a) Observation 2;
(b) Observations 3 and 4; and (c) Observation 5.  Power-law fits are shown for (a)
and (c).  For (b), the power spectrum is fitted using a power-law plus cutoff
power-law model.\label{fig7}}
\end{figure}

\begin{figure} \figurenum{8} \epsscale{1.0} \plotone{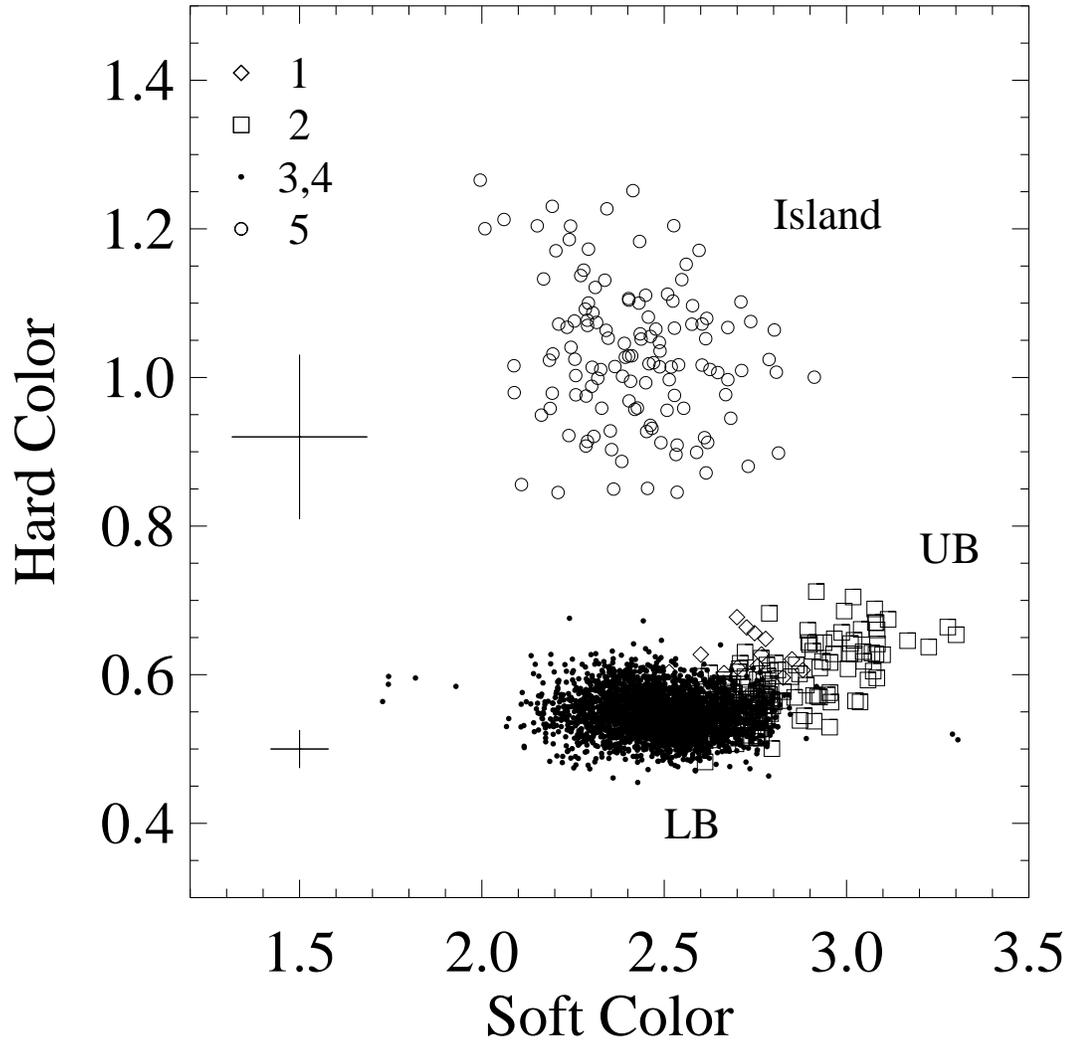}
\caption{Color-color diagram.  The soft color is the ratio between the 
3.5-6.4~keV and the 2.1-3.5~keV count rates, and the hard color is the ratio between 
the 9.8-20~keV and the 6.4-9.8~keV count rates.  For observations 1, 2, 3 and 4 
(diamonds, squares and filled circles), each point corresponds to 
16~s of data, and, for observation 5 (open circles), each point corresponds to 
128~s of data.  Sample error bars are shown on the left side of the figure.\label{fig8}}
\end{figure}

\begin{figure} \figurenum{9} \epsscale{1.0} \plotone{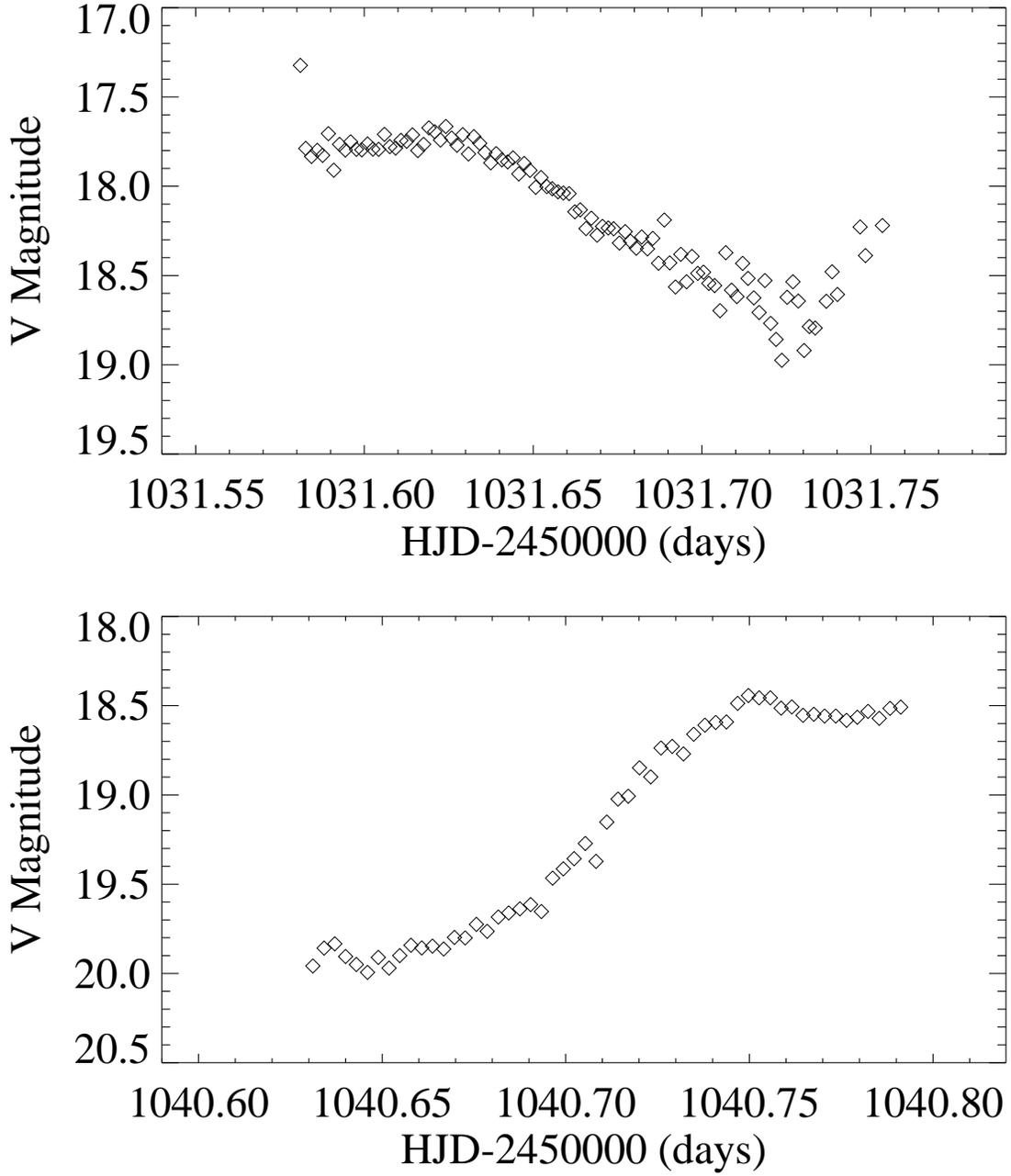}
\caption{The V-band light curves for the August 6 (top) and August 15 
(bottom) observations.\label{fig9}}
\end{figure}

\clearpage

%TABLES

\clearpage

\begin{deluxetable}{cccccc}
\footnotesize
\tablecaption{RXTE Observations and Burst Detections \label{tbl-1}}
\tablewidth{0pt}
\tablehead{\colhead{Observation} & \colhead{Date\tablenotemark{a}} & \colhead{Exposure Time (s)\tablenotemark{b}} & \colhead{Non-Burst Flux\tablenotemark{c}} & \colhead{Burst Start Time\tablenotemark{a}} & \colhead{Burst Stop Time\tablenotemark{a}}}
\startdata
1 & June 27 & 208 & $2.08\times 10^{-9}$ & June 27.98471 & June 27.98693\nl
\hline
2 & June 29 & 3392 & $1.82\times 10^{-9}$ & June 29.68425 & June 29.68527\nl
\hline
3 & July 14 & 19808 & $1.10\times 10^{-9}$ & July 14.45280 & July 14.45310\nl
 & & & & July 14.50451 & July 14.50487\nl
\hline
4 & July 22 & 25888 & $1.10\times 10^{-9}$ & July 22.23787 & July 22.23897\nl
 & & & & July 22.50956 & July 22.51042\nl
\hline
5 & August 15 & 16064 & $7.30\times 10^{-11}$ & No Bursts Detected & \nl
\tablenotetext{a}{1998 UT}
\tablenotetext{b}{This is the non-burst exposure time.}
\tablenotetext{c}{1.5--12 keV flux in erg cm$^{-2}$ s$^{-1}$}
\enddata
\end{deluxetable}

\begin{deluxetable}{lccc}
\footnotesize
\tablecaption{Optical Observations of XTE~J2123--058 \label{tbl-2}}
\tablewidth{0pt}
\tablehead{\colhead{Date (UT 1998)\tablenotemark{a}} & \colhead{Telescope} &
\colhead{Duration (hr)} & \colhead{Exposure Time (s)}}
\startdata
June 30.332 & MDM 1.3m & 2.74 & 30-300\nl
July 1.280 & MDM 2.4m & 3.00 & 30\nl
July 1.355 & MDM 1.3m & 0.24 & 120\nl
July 2.314 & MDM 1.3m & 1.90 & 30\nl
July 2.396 & MDM 2.4m & 1.83 & 30-60\nl
July 3.410 & MDM 2.4m & 1.61 & 30\nl
July 4.361 & MDM 1.3m & 2.85 & 30\nl
July 15.253 & KPNO 0.9m & 5.45 & 60\nl
July 16.255 & KPNO 0.9m & 4.70 & 60\nl
August 6.075 & CTIO 0.9m & 3.90 & 120\nl
August 15.124 & CITO 0.9m & 3.92 & 120\nl
September 20.671 & MDM 1.3m & 1.00 & 600\nl
\tablenotetext{a}{The date at the start of the XTE~J2123--058 observations.}
\enddata
\end{deluxetable}

\begin{deluxetable}{lccccc}
\footnotesize
\tablecaption{XTE~J2123--058 Spectral Results\tablenotemark{a} \label{tbl-3}}
\tablewidth{0pt}
\tablehead{\colhead{ } & \colhead{Obs. 1} & \colhead{Obs. 2} & \colhead{Obs. 3} & \colhead{Obs. 4} & \colhead{Obs. 5}}
\startdata
Comptonization\nl
kT (keV) & $3.03\pm 0.15$ & $2.67\pm 0.04$ & $2.86\pm 0.03$ & $2.74\pm 0.03$ & $27^{+530}_{-27}$\nl
$\tau$ & $11.54\pm 0.68$ & $12.56\pm 0.26$ & $10.97\pm 0.18$ & $11.47\pm 0.18$ & $3^{+40}_{-3}$\nl
Flux\tablenotemark{c} & 206 & 173 & 99 & 100 & 7.6\nl
N$_{\rm H}$ ($10^{22}$~cm$^{-2}$) & $1.14\pm 0.39$ & $0.98\pm 0.17$ & $0.79\pm 0.13$ & $0.88\pm 0.13$ & $0.00^{+0.05}_{-0.00}$\nl
$\chi^{2}/\nu$ & 40/44 & 46/44 & 65/44 & 52/44 & 36/44\nl
\hline
Comptonization\tablenotemark{b}\nl
kT (keV) & $2.77\pm 0.08$ & $2.54\pm 0.02$ & $2.73\pm 0.02$ & $2.62\pm 0.02$ & $23^{+239}_{-23}$\nl
$\tau$ & $13.23\pm 0.38$ & $13.90\pm 0.14$ & $11.84\pm 0.10$ & $12.47\pm 0.10$ & $4^{+22}_{-4}$\nl
Flux\tablenotemark{c} & 194 & 145 & 95 & 95 & 7.6\nl
$\chi^{2}/\nu$ & 48/45 & 77/45 & 97/45 & 92/45 & 37/45\nl
\hline
Blackbody plus Power-law\nl
kT (keV) & $1.86\pm 0.12$ & $1.94\pm 0.03$ & $1.93\pm 0.03$ & $1.91\pm 0.02$ & $0.67^{+0.06}_{-0.16}$\nl
R$_{BB}$ (in km for $d$=10~kpc) & $6.6\pm 0.9$ & $6.3\pm 0.2$ & $4.36\pm 0.13$ & $4.65\pm 0.12$ & $6\pm 4$\nl
$\Gamma$ & $2.49\pm 0.15$ & $2.95\pm 0.07$ & $3.03\pm 0.04$ & $3.06\pm 0.05$ & $1.85\pm 0.07$\nl
Flux$_{PL}$\tablenotemark{c} & 147 & 123 & 77 & 75 & 7.4\nl
N$_{\rm H}$~($10^{22}$~cm$^{-2}$)& $1.33^{+0.73}_{-0.77}$ & $2.57\pm 0.32$ & $2.35\pm 0.23$ & $2.50\pm 0.23$ & $0.0^{+0.8}_{-0.0}$\nl
$\chi^{2}/\nu$ & 35/43 & 44/43 & 60/43 & 45/43 & 26/43\nl
\hline
Blackbody plus Power-law\tablenotemark{b}\nl
kT (keV) & $1.72\pm 0.06$ & $1.756\pm 0.017$ & $1.742\pm 0.015$ & $1.739\pm 0.013$ & $0.65^{+0.06}_{-0.12}$\nl
R$_{BB}$ (in km for $d$=10~kpc) & $8.0\pm 0.6$ & $7.9\pm 0.2$ & $5.59\pm 0.11$ & $5.90\pm 0.09$ & $6\pm 2$\nl
$\Gamma$ & $2.28\pm 0.05$ & $2.51\pm 0.03$ & $2.663\pm 0.017$ & $2.675\pm 0.016$ & $1.85\pm 0.05$\nl
Flux$_{PL}$\tablenotemark{c} & 131 & 96 & 62 & 58 & 7.4\nl
$\chi^{2}/\nu$ & 38/44 & 103/44 & 154/44 & 149/44 & 26/44\nl
\tablenotetext{a}{The errors are 68\% confidence.}
\tablenotetext{b}{Column density fixed to $5.73\times 10^{20}$~cm$^{-2}$.}
\tablenotetext{c}{The unabsorbed 2.5-20~keV flux is in units of $10^{-11}$~erg~cm$^{-2}$~s$^{-1}$.}
\enddata
\end{deluxetable}

\begin{deluxetable}{ccccccc}
\footnotesize
\tablecaption{Fit Parameters for Power Spectra\tablenotemark{a}\label{tbl-4}}
\tablewidth{0pt}
\tablehead{ & & \multicolumn{2}{c}{Power-law} & \multicolumn{3}{c}{Cutoff Power-law}}
%\tablehead{\colhead{Observation} & \colhead{N\tablenotemark{b}} 
%& \colhead{$\alpha_{1}$} & \colhead{RMS Amplitude\tablenotemark{c}} & \colhead{$\alpha_{2}$} 
%& \colhead{$\nu_{c}$~(Hz)} & \colhead{RMS amplitude\tablenotemark{c}}}
\startdata
Observation & N\tablenotemark{b} & $\alpha_{1}$ & RMS Amplitude\tablenotemark{c} & $\alpha_{2}$ & $\nu_{c}$~(Hz) & RMS amplitude\tablenotemark{c}\nl
\hline
2 & 15 & $-1.10\pm 0.05$ & $2.84\pm 0.26$~\% & 0.6\tablenotemark{d} & 27\tablenotemark{d} & $<1$~\%\tablenotemark{e}\nl
3 & 143 & $-1.125\pm 0.033$ & $2.43\pm 0.15$~\% & $0.85\pm 0.44$ & $27\pm 10$ & $5.5\pm 0.4$~\%\nl
4 & 183 & $-1.433\pm 0.021$ & $2.25\pm 0.05$~\% & $0.64\pm 0.37$ & $21\pm 8$ & $4.4\pm 0.4$~\%\nl
5 & 117 & $-0.44\pm 0.06$ & $21.2\pm 3.5$~\% & 0.5 & 7 & $<21$~\%\tablenotemark{e}\nl
\tablenotetext{a}{The errors are 68\% confidence.}
\tablenotetext{b}{The number of 128 s power spectra.}
\tablenotetext{c}{In a frequency band from 0.01 to 100 Hz.}
\tablenotetext{d}{Fixed.}
\tablenotetext{e}{90\% confidence upper limits.}
\enddata
\end{deluxetable}

\begin{deluxetable}{lcccccc}
\footnotesize
\tablecaption{X-Ray and V-Band Fluxes \label{tbl-5}}
\tablewidth{0pt}
\tablehead{\colhead{Dates} & \colhead{$F_{x}$\tablenotemark{a}} 
& \colhead{V($\phi = 0.5$)} & \colhead{V($\phi = 0.0$)} & \colhead{Amplitude\tablenotemark{b}} & 
\colhead{$F_{\rm V,D\it}$\tablenotemark{c,d}} & \colhead{$F_{\rm V,OC\it}$\tablenotemark{c,d}}}
\startdata
July 1-16 & $9.6\times 10^{-10}$ & 16.90 & 17.65 & 0.75 & 4.13 & 4.16\nl
August 6 & $4.5\times 10^{-10}$ & 17.74 & 18.66 & 0.92 & 1.60 & 2.20\nl
August 15 & $7.30\times 10^{-11}$ & 18.49 & 19.98 & 1.49 & 0.442 & 1.44\nl
\tablenotetext{a}{The 1.5-12~keV flux is in units of erg~cm$^{-2}$~s$^{-1}$.}
\tablenotetext{b}{The optical modulation amplitude is in V magnitudes.}
\tablenotetext{c}{Inferred flux densities at $5.455\times 10^{14}$~Hz in units of $10^{-27}$
erg~cm$^{-2}$~s$^{-1}$~Hz$^{-1}$.}  
\tablenotetext{d}{A$_{V}$=0.3 is used to correct for extinction.}
\enddata
\end{deluxetable}

\end{document}